# Viscous fluid holographic inflation

E. Elizalde[1,2,3] and A. V. Timoshkin[2,3,4]

[1]*Institute of Space Sciences, National Higher Research Council of Spain (ICE/CSIC and IEEC) Campus UAB, C/ Can Magrans, s/n, 08193 Bellaterra (Barcelona), Spain*
[2]*Tomsk State Pedagogical University, Kievskaja Street, 60, 634061 Tomsk, Russia*
[3]*Tomsk State University of Control Systems and Radio electronics, Lenin Avenue, 36, 634050 Tomsk, Russia*
[4]*National Research Tomsk State University, Lenin Avenue, 36, 634050 Tomsk , Russia*

**Abstract**

A model of inflation produced by a viscous fluid is investigated and its compatibility with the holographic principle at the very early universe (as recently formulated for the holographic universe with a holographic cut-off radius) is demonstrated. Specifically, ensuing from the model, the corresponding scale factor and infrared cut-off are analytically calculated, which are taken to be the particle and future event horizon for inflation, respectively. Using them, the energy conservation law, in the holographic point of view, is obtained. In this way, total equivalence of viscous fluid inflation, with the specific cut-off of Nojiri and Odintsov, and holographic inflation is proven.

## 1. Introduction

The origin of the holographic principle is associated with the thermodynamics of black holes and string theory [1-4]. Mainly, the researches on the application of the holographic principle have been performed in late-time cosmology. In Ref. [5] a model of generalized holographic dark energy for an accelerating universe was discussed, in which the infrared cut-off is identified with a certain combination of the Friedmann-Lamaître-Robertson-Walker (FLRW) parameters. In Ref. [6], a model for holographic dark energy was proposed, following the idea that the short distance cut-off may be related with the infrared cut-off.

The holographic dark energy model is one of the existing models for quantum gravity. This possibility, based on the holographic principle, is proposed in Ref. [6] (see Ref [7] for a review); the holographic energy density is proportional to the inverse infrared cut-off $L_{IR}$ squared, namely,

$$\rho = \frac{3c^2}{k^2 L_{IR}^2}, \qquad (1)$$

where $k^2$ is the gravitational constant and $c$ is also a constant.

The most general holographic dark energy model was proposed by Nojiri and Odintsov in Ref. [8]. This holographic dark energy model describes large number of particular cases like Ricci dark energy, etc. Furthermore, it was recently shown that it may be generalized to holographic inflation (see Ref. [9]).

The holographic description of dark energy is of interest from a phenomenological point of view [10-12], since it can be connected to observations [13, 14]. Different versions of the cut-off, corresponding to various generalized holographic dark energies, have been considered in Refs. [15-21].

We should make clear that there is another previous definition of holographic inflation related with string theory and AdS/CFT duality [22-24], and that one should not at all mix these two different approaches to holography. Both terms have become kind of standard in the literature by now, but in this paper we use the term holographic inflation in the above defined sense, and not in the one of string theory. In short, there are two different approaches being ours not related with string theory holography.

In the present paper, we develop an investigation of a model of holographic inflation produced by a physical fluid, in which its viscosity plays an important role. In the inflationary scenario we will focus on the very initial stages of our universe. For various viscous fluid models, with constant and non-constant equation of state (EoS) parameters, we discuss the corresponding energy conservation law in terms of holographic inflation. And we establish the equivalence between our viscous fluid inflation and holographic inflation, within the specific cut-off choice of Nojiri and Odintsov [5].

## 2. Holographic inflation

The holographic principle at the very early universe has been discussed in Ref. [9] in some detail. Here, we will investigate holographic inflation as produced by a viscous fluid in a flat FLRW universe, with metric

$$ds^2 = -dt^2 + a^2(t) \sum_{i=1,2,3} \left(dx^i\right)^2, \qquad (2)$$

where $a(t)$ is the scale factor.

In the inflation period the first Friedmann equation reads

$$H^2 = \frac{k^2}{3} \rho_{\text{inf}}, \qquad (3)$$

where $\rho_{\text{inf}}$ is the energy density of the fluid that drives inflation. Its source can be, for example, a scalar field or modified gravity. We neglect the contribution from matter and radiation. Here $H(t) = \frac{\dot{a}(t)}{a(t)}$ is the varying Hubble parameter. In this work we consider that inflation has a holographic origin, namely that its source is the holographic energy density. Hence, imposing that $\rho$ within (1) is $\rho_{\text{inf}}$, the Friedmann equation for an expanding universe becomes simply

$$H = \frac{c}{L_{IR}}, \qquad (4)$$

where $c$ is a positive constant, since the expanding universe model is considered. As the infrared cut-off is related to causality, it must determine the form of the horizon.

As always, $H^{-1}$ corresponds to the radius of the cosmological horizon, which gives the limit of the causal relations from the viewpoint of the classical geometry, namely any information beyond the horizon cannot affect us, at least at the time when we measure $H$. The original idea of the holographic energy paradigm comes from identifying the classical horizon radius with the infrared cutoff in the quantum field theory, since we neglect any contribution coming from energy scales smaller than the cutoff scale. This implies that we may relate, as we will see later, the horizon radius $H^{-1}$ with the infrared cut-off, since the information in energy scales larger than the ultraviolet cutoff scale is irrelevant.

The simplest choice is that $L_{IR}$ should be exactly the Hubble radius, which however, cannot be used at late-times applications, since it cannot lead to an accelerating universe [18], and this is the case for the next guess, namely the particle horizon. Hence, one can use the future event horizon [6], the age of the universe or the conformal time [19, 20], the inverse square root of the Ricci curvature [21], a combination of Ricci, Gauss-Bonnet invariants, etc.

We thus consider the universe as a whole and use the cut-off radius of the horizon and its generalizations. This terminology is standard in all papers starting from M. Li [6] etc. As already mentioned above, there is another sort of holographic inflation related with string theory and AdS/CFT duality [22-24] and one should not mix these two approaches to holography. Both terms have become kind of standard in the literature, but in this paper we use the term holographic inflation in the above sense, not in the one of string theory.

There are various possibilities for the choice of the infrared radius, $L_{IR}$. For two of them, we may consider $L_{IR}$ as the particle horizon, $L_p$, or the future event horizon, $L_f$, which are defined, respectively, by

$$L_p \equiv a\int_0^t \frac{dt}{a}, \quad L_f \equiv a\int_t^\infty \frac{dt}{a}. \tag{5}$$

Such a choice of infrared radius allows us to apply the holographic principle to describe inflation. In a general situation, $L_{IR}$ could be a function of both $L_p$ and $L_f$ [8, 11]. Note that various special cases for holographic cut-off choices correspond to different specific choices of the Nojiri-Odintsov cut-off.

### 3. Representation of viscous fluid models in holographic inflation

In this section, we will apply the holographic principle to the inflationary universe for the description of inhomogeneous viscous fluid models in a flat FLRW space-time. We shall describe inflation in terms of the particle horizon, $L_p$, or the future event horizon, $L_f$, and the bulk viscosity. Next, we will use the results of Ref. [25].

Let us write the inhomogeneous EoS as

$$p = \omega(\rho)\rho + \zeta(H), \tag{6}$$

with $\omega(\rho)$ the thermodynamic parameter and $\zeta(H)$ the viscosity. We choose the viscosity in the form

$$\zeta(H) = \exp(-H/H_*) f(H), \qquad (7)$$

where $H_* = H(t_*)$ is the value of the Hubble function at the end of inflation $t_*$.

We assume that the universe is filled up with the fluid, and that the energy density $\rho$ satisfies the energy conservation law, namely,

$$\dot{\rho} + 3H(p+\rho) = 0. \qquad (8)$$

Let us now consider various models for the fluid.

### 3.1 Fluid model with constant $\omega(\rho) = \omega_0$ and viscosity proportional to Hubble square $H^2$

In the initial stage inflation, when $H/H_* \ll 1$, we have $\zeta(H/H_* \ll 1) = f(H)$. We choose the function $f(H)$ as

$$f(H) = \frac{b_0}{k^2} H^n \qquad (9)$$

were $b_0$ is in general a dimensional constant. We assume basically a power law for the bulk viscosity, as is quite usual in macroscopic cosmological theory. Let us, first make the choice $n = 2$, corresponding to a heavier weight on the $H$ dependence in the early universe, when $H$ was large. Physically that means the influence from bulk viscosity was taken to be large at the beginning of the universe's evolution.

The Hubble function reads

$$H = \frac{2H_{in}}{3\omega_0 H_{in}(t - t_{in}) + 2}, \qquad (10)$$

with $H_{in} = H(t_{in})$, where $t_{in}$ is the starting time of inflation and $\omega_0$ a dimensionless parameter.

The scale factor takes the form

$$a(t) = a_0 \left[3\omega_0 H_{in}(t - t_{in}) + 2\right]^{2/3\omega_0}, \qquad (11)$$

where $a_0$ is an integration constant.

Further, we can calculate the future event horizon, as

$$L_f = a \int_t^\infty \frac{dt}{a} = \frac{2}{2 - 3\omega_0} H^{-1}, \qquad (12)$$

Provided $\omega_0 < \frac{2}{3}$. From the holographic viewpoint $H$ and its derivative $\dot{H}$ can be represented as the future event horizon $L_f$ [5]:

$$H = \frac{\dot{L}_f}{L_f} + \frac{1}{L_f}, \quad \dot{H} = \frac{\ddot{L}_f}{L_f} - \frac{\dot{L}_f^2}{L_f^2} - \frac{\dot{L}_f}{L_f^2}. \tag{13}$$

Thus, by using (13), the conservation law of the fluid in (8) in the holographic language can be rewritten as

$$2\left(\frac{\ddot{L}_f}{L_f} - \frac{\dot{L}_f^2}{L_f^2} - \frac{\dot{L}_f}{L_f^2}\right) + 3\omega_0\left(\frac{\dot{L}_f}{L_f} + \frac{1}{L_f}\right)^2 = 0. \tag{14}$$

Next, we obtain the corresponding EoS, which turns out to be

$$p = \frac{3\omega_0 + 1}{k^2}\left(\frac{\dot{L}_f + 1}{L_f}\right)^2. \tag{15}$$

To summarize, in this first model we have successfully investigated the application of the holographic principle in the initial stages of inflation.

### 3.2 Fluid model with constant $\omega(\rho) = -\frac{\rho}{\rho + \rho_*}$ and viscosity proportional to $H$

Similarly, to the previous section, we will now analyze this model in the asymptotic limit corresponding to the initial stages of inflation. We take here the function $f(H)$ to be proportional to $H$ (thus, $n = 1$ in Eq. 9)

$$f(H) = \frac{3H_{in}}{k^2}H. \tag{16}$$

Most often, one makes the choice of a linear dependence of the viscosity on $H$ [26]. From the physical point of view, that is a milder influence from the bulk viscosity at the beginning of inflation, and a more conventional form.

As we consider the initial stage of inflation, we can choose the energy density, $\rho_*$, as $\rho_* = \frac{3}{k^2}H_{in}^2$.

In this approximation, the Hubble function becomes [25]

$$H(t) = \left(\sqrt{\tau^2 + 1} - \tau\right)H_{in}, \tag{17}$$

where $\tau = \frac{3}{4}H_{in}(t - t_{in})$.

Now, we can calculate the scale factor, with the result

$$a(t) = a_0 \left[ \left( \sqrt{\tau^2 + 1} + \tau \right) e^{\tau \left( \sqrt{\tau^2+1} - \tau \right)} \right]^{\frac{2}{3}}. \tag{18}$$

In the particular case, when $\tau \to 0$ $(t \to t_{in})$, in the linear approximation, the expression (18) simplifies, and one obtains

$$a(\tau \to 0) \approx a_0 \left( 1 + \frac{4}{3} \tau \right). \tag{19}$$

We can now describe the holographic inflation in terms of the particle horizon $L_p$, as

$$L_p \approx a \int_{t_{in}}^{t} \frac{dt}{a} = \left( 1 + \frac{4}{3} \tau \right) \ln \left( 1 + \frac{4}{3} \tau \right) H_{in}^{-1}. \tag{20}$$

From the holographic viewpoint, $H$ can be represented as the particle horizon $L_p$ [5]

$$H = \frac{\dot{L}_P - 1}{L_p}, \quad \dot{H} = \frac{\ddot{L}_p}{L_p} - \frac{\dot{L}_p^2}{L_p^2} + \frac{\dot{L}_p}{L_p^2}. \tag{21}$$

Consequently, the energy conservation law for the fluid (8) can be rewritten as

$$\frac{2}{3} \left( \frac{\ddot{L}_p}{L_p} - \frac{\dot{L}_p^2}{L_p^2} + \frac{\dot{L}_p}{L_p^2} \right) + \left( \frac{\dot{L}_P - 1}{L_p} \right)^2 \frac{\rho_*}{\frac{3}{k^2} \left( \frac{\dot{L}_P - 1}{L_p} \right)^2 + \rho_*} = 0. \tag{22}$$

Summing up, we note that we have just obtained the reconstruction of the arbitrary viscous fluid for inflation as a generalized holographic energy.

### 3.3 Quasi-de Sitter expansion for inflation

Finally, we consider a non-viscous model for the fluid, which describes quasi-de Sitter inflation. The thermodynamic parameter has the form [25]

$$\omega(\rho) = -1 + b_1 \rho^{\frac{1}{2}} - b_2 \rho^{-\frac{1}{2}}, \tag{23}$$

where $b_1$ and $b_2$ are positive dimensional constants.

Let us set the constants $b_1$ and $b_2$ equal to

$$b_1 = \frac{1}{\sqrt{3\theta}}, b_2 = \delta b_1, \tag{24}$$

respectively, where $\theta = \frac{3}{k^2}$ and $\delta$ is a positive dimensional constant.

The solution of the Eq. (8) yields the Hubble parameter in the form [25]

$$H = \left[\frac{3}{\sqrt{\theta}}(t-t_{in}) + \frac{1}{H_{in}^2}\right]^{-\frac{1}{2}}. \tag{25}$$

The scale factor is given by the expression

$$a(t) = a_0 e^{\frac{2}{3}\sqrt{\theta}\sqrt{\frac{3}{\sqrt{\theta}}(t-t_{in}) + \frac{1}{H_{in}^2}}}. \tag{26}$$

Here a future horizon can be defined by

$$L_f = H^{-1} + \frac{3}{2\sqrt{\theta}}. \tag{27}$$

Using the holographic language, we may rewrite the continuity equation for the energy as follows

$$\dot{L}_f - \frac{3}{4\sqrt{\theta}\left(L_f - \frac{3}{2}\sqrt{\theta}\right)} = 0. \tag{28}$$

To conclude this section, in the three examples above, we have explicitly exhibited the possibility of application of the holographic principle for the description of the inflationary universe, within the quite standard setting of viscous fluid cosmology.

## 4. Conclusion

In this work we have shown the possibility to apply the holographic principle to the very early stages of the universe evolution. Three specific cosmological models for describing a holographic inflation scenario with the use of an inhomogeneous EoS, relying, in special, on the viscosity property of the fluid, have been considered. We have paid attention to the very initial stage of the inflation. For each model we have analytically calculated the scale factor, the particle or future event horizons. One should notice that a change in sign in the viscosity term in Eq. (6) may lead to the appearance of future singularities (see, e.g., [27]). For our discussion here we preferred not to deal with this issue, what we expect to do elsewhere.

Specifically, we have been able to rewrite the continuity equation for the energy density of the viscous fluid in the holographic language. The equivalence in describing inflation by means of a viscous fluid versus holographic inflation with the specific cut-off of Nojiri and Odintsov has been shown, in particular, with three specific examples of the application of the holographic description of inflation.

This theory can be extended to the case of two coupled fluids. All calculations are similar to those presented here. Note that, applying the above method, one can visualize inflation from modified gravity [28] as holographic inflation, too.

**Acknowledgments**

This work was partially supported by MINECO (Spain), FIS2016-76363-P, by the CPAN Consolider Ingenio 2010 Project, and by AGAUR (Catalan Government), project 2017-SGR-247. EE also acknowledges the kind hospitality of Iver Brevik at NTNU (Norway), where this paper was finished, and support from the NRC, project No. 250346. We thank R. Baier for useful comments.

**References**

[1] G. 't Hooft, Salamfest 1993: 0284-296 [gr-qc].

[2] L. Susskind, J. Math. Phys. **36**, 6377 (1995).

[3] E. Witten, Adv. Theor. Math. Phys. **2**, 253 (1998).

[4] R. Bousso, Rev. Mod. Phys. **74**, 825 (2002).

[5] S. Nojiri, S. D. Odintsov, arXiv:1703.06372 [hep-th], (2017).

[6] M. Li, Phys. Lett. B **603**, 1 (2004).

[7] S. Wang, Y. Wang and M. Li, Phys. Rept. **696**, 1 (2017).

[8] S. Nojiri and S. D. Odintsov, Gen. Rel. Grav. **38**, 1285 (2006).

[9] S. Nojiri, S. D. Odintsov and E. N. Saridakis, arXiv:1904.01345v1 [gr-qc] (2019).

[10] D. Pavon and W. Zimdahl, Phys. Lett. B **628**, 206 (2005).

[11] E. Elizalde, S. Nojiri, S. D. Odintsov and P. Wang, Phys. Rev. D **71**, 103504 (2005).

[12] X. Zhang and F. Q. Wu, Phys. Rev. D **72**, 043524 (2005).

[13] M. Li, X. D. Li, S. Wang and X. Wang, JCAP **0906**, 036 (2009).

[14] J. Lu, E. N. Saridakis, M. R. Setare and L. Xu, JCAP **1003**, 031 (2010).

[15] M. Khurshudyan, Astrophys. Space Sci. **361**, no. 12, 392 (2016).

[16] L. Zhang, P. Wu and H. Yu, Eur. Phys. J. C **71,** 1588 (2011).

[17] E. N. Saridakis, Phys. Lett. B **661**, 335 (2008).

[18] S. D. H. Hsu, Phys. Lett. B **594**, 13 (2004).

[19] R. G. Cai, Phys. Lett. B **660**, 228 (2007).

[20] H. Wei and R. G. Cai, Phys. Lett. B **660**, 113 (2008).

[21] C. Cao, X. Chen and Y. G. Shen, Phys. Rev. D **79**, 043511 (2009).

[22] J. Maldacena, JHEP **0305**, 013 (2003).

[23] Paul McFadden, K. Skenderis, J. Phys.: Conf. Ser. **222**, 012007 (2010).

[24] A. Bzowski, Paul McFadden, K. Skenderis, JHEP **1304**, 047 (2013).

[25] I. Brevik and A. V. Timoshkin, JETF **122**, 679 (2016).

[26] K. Bamba, S. Nojiri, S. D. Odintsov and D. Saez-Gomez, Phys. Rev. D **90**, 124061 (2014).

[27] B.D. Normann and I. Brevik, Entropy **18**, 215 (2016).

[28] S. Nojiri, S. D. Odintsov and V. K. Oikonomou, Phys. Rep. **692**, 1 (2017).